\documentclass[conference]{IEEEtran}
\IEEEoverridecommandlockouts
\usepackage{cite}
\usepackage{pdfpages}
\usepackage{amsmath,amssymb,amsfonts}
\usepackage{algorithmic}
\usepackage{graphicx}
\usepackage{textcomp}
\usepackage{xcolor}
\usepackage{url}
\def\BibTeX{{\rm B\kern-.05em{\sc i\kern-.025em b}\kern-.08em
    T\kern-.1667em\lower.7ex\hbox{E}\kern-.125emX}}

\makeatletter
\newcommand{\linebreakand}{%
  \end{@IEEEauthorhalign}
  \hfill\mbox{}\par
  \mbox{}\hfill\begin{@IEEEauthorhalign}
}
\makeatother

\makeatletter
\def\ps@IEEEtitlepagestyle{%
\def\@oddfoot{\mycopyrightnotice}%
\def\@evenfoot{}%
}
\def\mycopyrightnotice{%
{\footnotesize 978-1-6654-3326-6/21/\$31.00~\copyright~2021 IEEE\hfill} 
\gdef\mycopyrightnotice{}
}

\begin{document}
\null%
\includepdf[pages={1}]{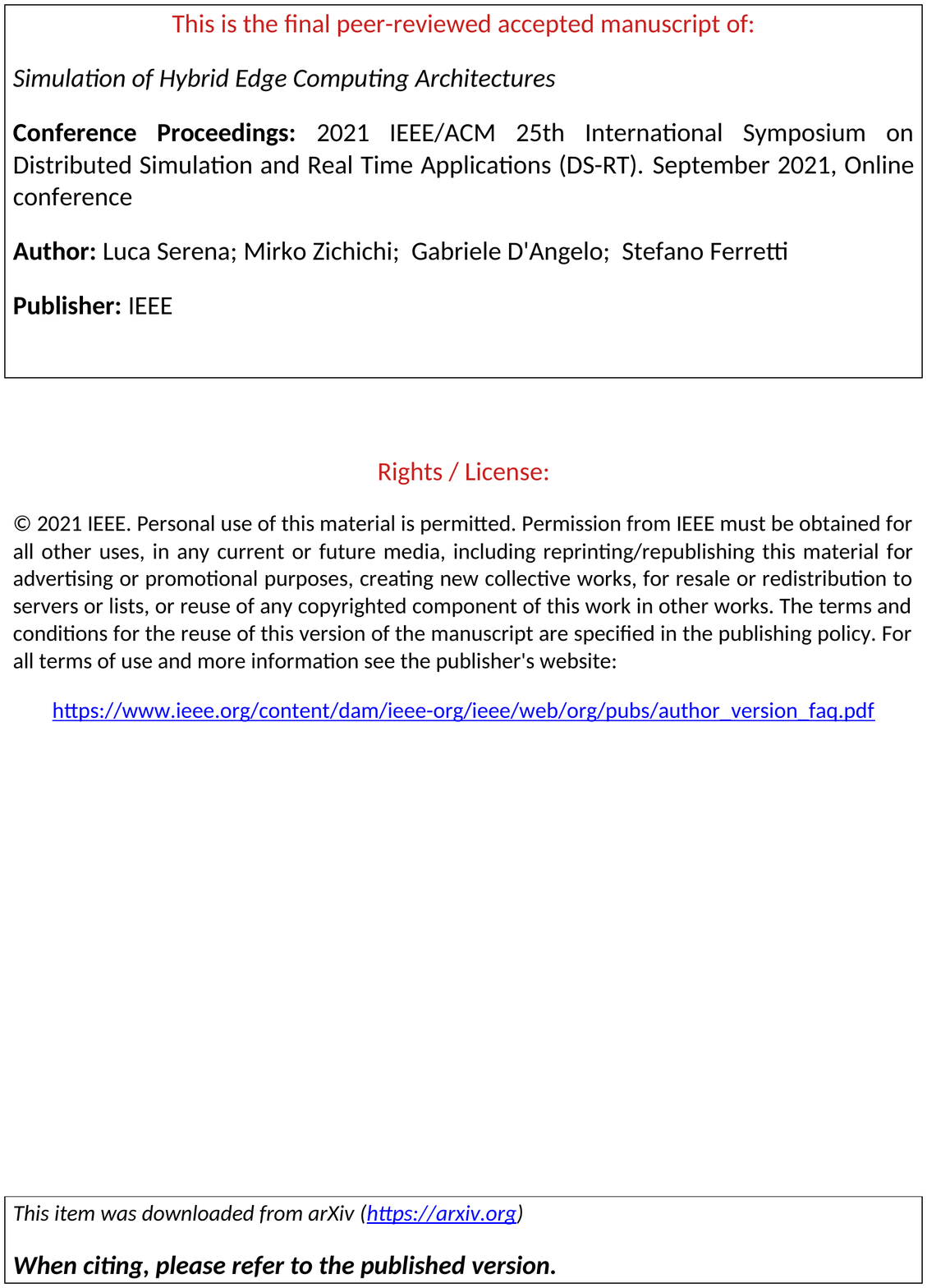}

\title{Simulation of Hybrid Edge Computing Architectures}

\author{\IEEEauthorblockN{Luca Serena}
\IEEEauthorblockA{\textit{Department of Computer Science and Engineering}\\
\textit{University of Bologna}\\
Bologna, Italy\\
luca.serena2@unibo.it}
\and
\IEEEauthorblockN{Mirko Zichichi}
\IEEEauthorblockA{\textit{Ontology Engineering Group} \\
\textit{Universidad Politécnica de Madrid}\\
Madrid, Spain\\
mirko.zichichi@upm.es}
\linebreakand
\IEEEauthorblockN{Gabriele D'Angelo}
\IEEEauthorblockA{\textit{Department of Computer Science and Engineering} \\
\textit{University of Bologna}\\
Cesena, Italy\\
g.dangelo@unibo.it}
\and
\IEEEauthorblockN{Stefano Ferretti}
\IEEEauthorblockA{\textit{Department of Pure and Applied Sciences}\\
\textit{University of Urbino ``Carlo Bo"}\\
Urbino, Italy\\
stefano.ferretti@uniurb.it}
}

\maketitle

\begin{abstract}
Dealing with a growing amount of data is a crucial challenge for the future of information and communication technologies. More and more devices are expected to transfer data through the Internet, therefore new solutions have to be designed in order to guarantee low latency and efficient traffic management. In this paper, we propose a solution that combines the edge computing paradigm with a decentralized communication approach based on Peer-to-Peer (P2P). According to the proposed scheme, participants to the system are employed to relay messages of other devices, so as to reach a destination (usually a server at the edge of the network) even in absence of an Internet connection. This approach can be useful in dynamic and crowded environments, allowing the system to outsource part of the traffic management from the Cloud servers to end-devices. To evaluate our proposal, we carry out some experiments with the help of LUNES, an open source discrete events simulator specifically designed for distributed environments. In our simulations, we tested several system configurations in order to understand the impact of the algorithms involved in the data dissemination and some possible network arrangements.
\end{abstract}


\section{Introduction}
We are living in an era in which digital services are constantly transformed and revised. All the tools that people use are now digital, producing some kind of data that is not necessarily stored in a local storage, but that often needs to be uploaded to distributed systems, through some communication means. With the rise of the Internet of Things (IoT), an increasingly higher number of devices is expected to join the Internet in a near future, interacting with some form of Cloud or decentralized platforms~\cite{zichichi2020framework}. In order to manage the growing amount of traffic different novel technological solutions are being proposed. Among them, 5G stands out, which is capable of offering Internet access to a significantly higher number of mobile devices with an improved efficiency~\cite{akpakwu2017survey}.

In this context, smart cities and smart shires~\cite{ferretti2016smart} are supposed to emerge, with the employment of hybrid physical-digital and intelligent infrastructures that use data-driven technologies to adapt to changes in the physical environment~\cite{chourabi2012understanding}. However, the growth of data exchanges between devices needs to be managed not only from a network infrastructure point of view, but also from the perspective of Cloud platforms, in order to avoid an overload of requests to the servers and the resulting increased latencies or service unavailability~\cite{kiss2018deployment}.

Edge computing thus emerges as a paradigm for improving the efficiency of the content delivery, by decentralizing the management of the system and bringing computation and data storage in locations geographically closer to the users. With an edge computing approach, most of the activities usually performed by computers in data centers are now carried out by some edge servers, situated in the vicinity of the end-users. This strategy can lead us to various benefits, such as minimizing the network traffic, decreasing the latency time, real-time execution, event-driven developments and efficient deployments~\cite{aslanpour2020performance}.

It is worth mentioning that this whole scenario is strictly related to a novel concept of Internet of People (IoP), a paradigm devoted to putting individuals and their personal devices at the heart of data management design~\cite{conti2018internet}. Smartphones and personal IoT devices play an active role in data management by autonomously building and configuring the services their users need, instead of delegating these tasks to centralized remote platforms. In the IoP paradigm, the frontiers of computing applications, data and services are pushed away from centralized servers to edge and end devices. It also: (i) makes it possible to design new crowdsourced and socially empowered architectures; (ii) shifts trust towards cryptographic techniques and network consensus mechanisms; (iii) allocates or delegates computation, synchronization, and storage to other edge devices; (iv) enables autonomous decision-making at the network frontiers; and (v) exploits physical proximity to create peer-to-peer (P2P) systems and content distribution networks~\cite{garcia2015edge}. All these features are able to foster greater efficiency in the communication and distribution of information between users (logically or physically) close to each other and, above all, are able to reduce the centralization of current online platforms.

This whole scenario depicts a wide set of possible architectural solutions for the deployment of scalable and effective distributed services. These solutions must be adapted to the specific use-case, taking into account the location, geographical characteristics, available infrastructures, possible impediments and constraints. All these aspects can influence the way the involved digital actors can interact, the underlying communication technologies and even the topology of the resulting interaction networks~\cite{ccnc2020,gda-simpat-iot,Ferretti2013481}. To sum up, it is not possible to foresee a single solution that fits all the requirements. Rather, there is the need to devise configurable and adaptable strategies for the distribution of services. This means that there is a strong need for tools that allow to evaluate, during the design phase or at runtime, complex distributed systems such as those related to edge computing and P2P ones.

In this paper, we show how such a kind of ``ex-ante'' evaluation process can be accomplished, through the use of a simulator called LUNES (i.e.~Large Unstructured Network Simulator) \cite{pads}. In particular, we study a distributed architectural solution that merges P2P interactions with the edge computing paradigm. According to this scheme, end-node devices are exploited for relaying messages, which eventually will have to be delivered to one of the edge nodes. 
We carried out some experiments in order to evaluate how such a system could be designed. In our model, we use a multi-layer graph, populated by two types of nodes: the end-nodes (i.e.~the devices of the users) and the edge nodes (i.e. ~computers that have the function of decentralized servers). Simulated nodes are located in a 2-dimensional space, and they can communicate with the peers that are sufficiently close. The simulation testbed is thought do be dynamic, with end-nodes being able to move along the grid (i.e.~a discrete space environment represented as a $N*N$ Cartesian plane composed of $N^2$ cells, each having as coordinates an integer ranging from $0$ to $N-1$). Simulation is divided into several time-steps, and in a single unit of time devices can move to an adjacent cell and nodes can forward a received message. 
Multiple design choices may have a significant impact on the efficiency of the system, such as the number of the edge nodes, their disposition and the gossip protocol being used to disseminate messages.
This kind of design should be suitable for Smart Shires and Smart Cities environments, where several devices are thought to send and receive a significant amount of information and where therefore the communication mechanisms are of crucial importance.

Obtained results provide some important insights. First, they confirm that the use of simulation is an important means to perform ``what-if analyses'' and to study the possible performance of a designed system. Through the use of simulation, one can simply change some configurations of a system and to evaluate them, without the need for costly changes in a deployed system. For instance, in our simulations we varied the number of nodes, the percentage of mobile and fixed nodes, the position of the fixed nodes, the dissemination protocols used to propagate information in the system and the mobility algorithm followed by the mobile nodes. Second, results show that with the proposed solution a very big number of messages is sent through the network. However, by adopting the appropriate measures, it is possible to considerably improve the efficiency in terms of traffic minimization without compromising the time delivery and the successful communication rate achieved.

The remainder of this paper is organized as follows. Section 2 introduces some background and related work. Section 3 describes the design choices of the software tool and deals with the critical aspects of the implementation. Section 4 analyzes the results obtained by testing different system configurations. Finally, Section 5 provides some concluding remarks.

\section{Background}
In this section we introduce the background and related work that is necessary to properly describe the proposed architectural solution.

\subsection{Peer-to-Peer and Edge Computing}
Peer-to-Peer platforms are systems where several computers form an overlay network (usually running on top of the Internet) and manage communication and resources sharing without the presence of a central authority being involved. These types of applications were originally created for file sharing (e.g.~BitTorrent) and recently raised their popularity due to the advent of blockchains and cryptocurrencies~\cite{ccnc2020,gda-ppna-2021}. Communication in a P2P environment is a crucial issue and there are different solutions to ensure that two nodes can exchange data~\cite{serena2020implications}. Hybrid P2P systems might employ some servers for coordination, to whom peers could ask the IP addresses of the other peers owning a certain resource. Pure P2P architectures, instead, do not rely on any server and a dissemination protocol is employed~\cite{backx2002comparison}.

On the other hand, edge computing is a paradigm whose purpose is to bring computation closer to the end-users, by setting up several edge nodes, which are lightweight servers placed geographically as close as possible to the users~\cite{shi2016edge}. Decentralization of content storage and delivery may have multiple positive effects, such as:
\begin{itemize}
    \item \emph{Reduced latency} - the end-node devices turn to the closest server and thus the reduced distance leads to a smaller delay in communication, making real-time execution closer to accomplishment.
    \item \emph{Reducing data center workload} - also leading to a more sustainable energy consumption\cite{varghese2016challenges}.
    \item \emph{No single point of failure} - in case servers in a specific data center are temporarily not available for malfunction or maintenance, then the content still remains retrievable.
\end{itemize}
There have already been studies and proposals to combine P2P and edge computing paradigms, for example in \cite{karagiannis2019edge} a P2P communication approach among the edge nodes has been proposed. Other works, such as \cite{yadgar2019modeling} highlighted some similarity in management and structuring between P2P and edge computing, while in \cite{ko2018wireless} spatial modelling was used to investigate computing and communication latencies in an edge computing environment. 

\subsection{Mobility Algorithm}
The typical IoT deployments include applications for both static and mobile devices. In fact, devices emitting signals can either be in a fixed location (e.g.~a house appliance) or they can change their geographical location over time (e.g.~smartphones, cars, drones). Multiple schemes can be used in order to reproduce such movements, taking also into account the human behaviour that triggers such movements.
\begin{itemize}
    \item \emph{Static model} - The end-nodes (i.e.~devices) are situated in a random place in the grid and they will stay still throughout the full duration of the tests. 
    \item \emph{Random independent movements model} - At each time-step the nodes have a probability $p$ to move into an adjacent cell and probability $1-p$ to stay still.
    \item \emph{Random Waypoint model} - In this widely used movement model, a node is either stationary or in motion toward a certain location. Stationary nodes have a certain probability to activate, by choosing a random location of the grid as a destination. When a destination is picked, then the node begins to move toward that point with a given speed~\cite{hyytia2007random}. In our model, since the simulation steps represent small time-units, the nodes only move to an adjacent cell in one time-step.
    \item \emph{Community-based model} - This model is thought to represent groups of individuals that behave in a similar manner~\cite{vastardis2014enhanced}. When a stationary node $n$ activates by choosing a destination point, then also the other stationary nodes close to $n$ will head towards such a destination cell.
\end{itemize}

\subsection{Dissemination Algorithms}
In a P2P environment, for scalability reasons the nodes are directly in touch only with a bunch of peers, and they do not know the location of the other nodes. Thus, in the communication process the information is relayed multiple times among the participants of the system (i.e.~multi-hop), until the final destination is reached. In particular, in our use case the communication is wireless, hence, similarly to Bluetooth, only the devices within a certain range are reachable. To define the policy for messages dissemination in a P2P environment a gossip protocol is employed. Different types of algorithms can be implemented depending on the semantics of the system. In some networks achieving a very high coverage (i.e.~percentage of peers who receive the message) is fundamental, while other ones may be focused on traffic minimization or retention of anonymity. 
This issue is further exacerbated in an edge computing scenario, where multiple nodes can communicate through some ad-hoc or mesh networking solutions; hence via short range wireless communication means, e.g.~WiFi direct, bluetooth, LoRa. In this case, the communication overlay is formed through the communication range, i.e.~each node is considered as connected only with those nodes that are at a reachable distance, due to the used wireless communication technology.
In our work, we will consider the following dissemination algorithms:
\begin{itemize}
    \item \emph{Pure broadcast} - The message is forwarded to all the neighbors, except the forwarder. In this way we achieve the theoretical minimum time deliver and the maximum coverage, at the cost of an high amount of network traffic.
    \item \emph{Probabilistic Broadcast} - Given a forwarding parameter $p$, there is $p\%$ of chance that a node forwards the message to all the neighbors and $(100-p)\%$ that it does not send it to any other node.
    \item\emph{Reduced Range} - Since in wireless communication a signal is spread through air, then it is not possible to arbitrarily deliver the content to a limited set of receivers. Therefore, an alternative for traffic minimization is to reduce the power of the signal, thus reaching a lower number of peers. Given a parameter $p$, a signal is spread just to the $p\%$ of its normal power range $d$, thus reaching on average $(d-p)^2/d^2$ of the nodes with respect to the standard configuration.
    \item\emph{Directed Propagation} - Following the same principle of Reduced Range, the purpose is to reach fewer nodes with the propagation of the signal. In particular, the signal is propagated only toward certain directions. In this case some geographical information about the environment can be exploited.
\end{itemize}

All the dissemination schemes also need mechanisms to avoid infinite loops of messages, avoiding to forward already received data or setting a time-to-live for messages (i.e.~a message can be relayed just a certain amount of times, so at every hop a counter is decreased). Multiple gossip protocols exist other than the aforementioned ones, but some are impractical in such a context. For example, protocols which require the knowledge of the number of connections of the other nodes cannot be applied since in this scenario connections are fleeting, so it is unrealistic to base on information describing the connections state at a given time. Also forwarding messages to a limited number of peers is not applicable, since in our use case the signals propagate through air, thus messages are not manageable and routable.

The metrics used to evaluate the performance of such algorithms are the following:
\begin{itemize}
    \item \emph{Successful communication rate} - It indicates the percentage of times that a specific node was able to get in touch with a designated node, with respect to all attempts made.
    \item \emph{Messages sent} - It indicates the average number of messages being sent in the process of getting in touch with the recipient node. Both successful and unsuccessful attempts are counted.
    \item \emph{Delay} - It indicates the number of discrete time-steps needed on average to contact the recipient node. In our use cases, this metric also corresponds to the number of hops needed for the delivery.
\end{itemize}

\section{Large Unstructured Network Simulator (LUNES)}
LUNES is a time-stepped discrete event simulator for complex networks~\cite{gda-jpdc-2017}, which allows to simulate certain network protocols and to evaluate their efficiency. LUNES is implemented on top of ART\`IS/GAIA simulation middle\-wa\-re~\cite{bononi2005scalable}, which implements the primitives for communication among simulated entities and time management, also offering support for parallel and distributed execution~\cite{gda-simpat-2017}.

Scalability of simulations is one of the main issues that LUNES wants to solve, allowing it to run over {10\,000} simulated entities in a single machine. The nodes of the system are labelled with an integer ID and possibly with other state variables, which describe some of their features, thus enabling the modelling of multilayer and temporal graphs. Two functions of the simulator are particularly important for the execution: a first one that is triggered at each time-step for all the nodes of the simulation, performing actions if needed and a second one that is triggered every time a message is received.

LUNES was designed to be easily adaptable to various distributed environment configurations, allowing the users to model the protocols to be tested and the features of the simulated entities involved. In previous works, LUNES was employed to evaluate the impact of certain attacks on blockchains~\cite{serena2020implications} or to simulate the dissemination of game events in P2P Multiplayer Online Games~\cite{Ferretti2010}.

The peculiarity of the LUNES version used for the experiments reported in this paper, is that nodes have a geographical position and no fixed neighbors, and the communication is based on their physical distance. When a node is going to send a message, it scans a list of simulated entities delivering the message only to the nodes within a certain range (a parameter that fixes the communication distance is set). In order to reduce the complexity of the scan operation, at the beginning of each epoch, a list of potentially ``close enough'' nodes is created for all the simulated entities, and until the next epoch only those nodes are considered as potential receivers for a certain node. This version of the simulator is built according to a multi-level approach, with the time being divided into epochs (i.e.~an epoch is the fraction of the simulation where a single experiment is performed)~\cite{gda-simpat-iot}. At the first step of an epoch an applicant node is chosen and such peer will spread its request to the network. In the remaining time-steps messages are propagated through the network. An epoch must last a sufficient number of steps to guarantee that the delivery of a message, if possible, is carried out before the start of the following epoch.
The testbed of the simulator is a multilayer graph (whose details are explained in the following section) where the end-nodes follow a mobility model and relay the new received messages, while edge nodes are static and represent the end point of the communication: if an edge node is reached then the test is considered as successful.

\section{Performance Evaluation}
In this performance evaluation we investigate a hybrid edge-P2P system, in which the end-nodes try to get in touch with an edge node by relaying their requests to the other nodes of the system, until a destination point (i.e.~one of the edge nodes) is reached. 
In order to represent such a system, the nodes are disposed in a grid and therefore each participant is associated with a geographical position.  The system is represented by a multilayer graph, being the two layers a representation of this hybrid configuration of the distributed system, i.e.~the mobile P2P layer and the edge computing layer. As a consequence, there are two types of nodes: (i) 
peer nodes (end-users) and (ii) edge nodes. Each node can communicate with all the other nodes placed within a certain distance, and the end-users move along the grid during the simulation. Consequently, neighborhood relations are not stable, but they are subject to changes (more or less sudden depending on the mobility model) over time. 

Different factors can influence the efficiency of the communication:
\begin{itemize}
    \item \emph{Gossip protocol} - In this context we are mainly concerned with the amount of time that is necessary to deliver a message, since one the main goal of edge computing is to reduce the communication latency. Pure broadcast guarantees the fastest possible delivery, but other protocols can be used to minimize traffic, above all when nodes are situated in a crowded environment.
    \item \emph{Mobility model} - Depending on the specific application to be simulated, appropriate mobility models can be used to simulate the movements of the actors involved.
    \item \emph{Communication range} - In this model, thought for wireless communications, we assume that nodes can directly exchange data when they are situated within a certain distance. Changing such a parameter would have strong repercussions on the metrics: if the communication range is too short, nodes risk to struggle to get in touch with other peers, and the number of hops (and therefore time) before reaching an edge node might be high. On the other hand, if the range is too high, the P2P aspects of the model would become irrelevant, since end-users would tend to get immediately in touch with the edge nodes.
    \item \emph{Nodes density} - Similarly to the models with short-range communication distance, if the graph is sparsely populated then there is the risk that the requests do not reach their destination through the use of relays among peers. On the other hand, a very crowded environment could lead to an enormous amount of network traffic for relaying the message to the destination. This problem, however, could be easily solved by adopting a proper gossip protocol.
    \item \emph{Amount and position of edge nodes} - The more edge nodes in the network, the lower the latency for communication. Furthermore, an optimized and targeted position of the edge nodes can considerably reduce the time for delivery.
\end{itemize}
In the default configuration, used for the following tests, {10\,000} nodes populate a $1000$ x $1000$ grid, thus on average there is a node every $100$ cells. The communication radius is set to $40$, which means that on average a node reaches $50$ other nodes within a circular area of $40^2*\pi$ cells, unless it is positioned on the edges of the grid. Where it is not specified differently, pure broadcast and Random Waypoint are used respectively as the gossip protocol and the mobility algorithm. Furthermore, the time-to-live for messages has been set equal to 20, even though it was noticed that this value is widely abundant.

\subsection{Mobility model}
\begin{figure}
    \centering
    \includegraphics[width=.45\textwidth]{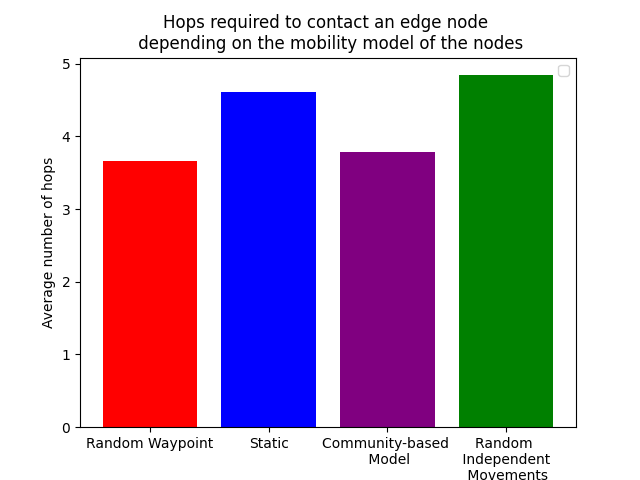}
    \caption{Average number of hops necessary to contact an edge node, depending on the mobility model of the nodes.}
    \label{mmod}
\end{figure}

Figure \ref{mmod} shows how different mobility models of nodes can influence the average number of hops needed to contact an edge node. In particular, with a static mobility setup, fewer messages are sent and consequently the delay for the delivery of the message is higher. This is actually due to the testbed chosen for the experiments. In our model, edge nodes are placed in an ``optimized position'' at the center of the Cartesian plane, so that the distance between a cell and an edge node is minimized. However, with such a configuration, end-nodes at the edge of the Cartesian plane are the most distant from the edge nodes. Furthermore, these nodes forward (on average) fewer messages than the other nodes, since the borders of the grid limit the usable surface where the signal is propagated (assuming that over the borders of the grid no node is placed).

Clearly enough, in the static configuration, the number of nodes placed at the edge of the Cartesian plane (i.e.~fewer than 40 cells from the border) is constant, whereas with Random Waypoint the moving nodes tend to stay more time at the center of the grid while reaching the destination passing through the shortest path. This is due to the fact that the simulation environment used as a testbed is not toroidal. As expected (and reported in the related literature about mobility models), we have noticed that in Random Waypoint, after an initial adjustment period, the number of nodes at the edge of the grid is between $400$ and $600$ (around 5\% of the total nodes), while in the static configuration around 15\% of the nodes is located in such a critical position. 

Similar remarks are possible for other algorithms: community-based model presents a percentage of peers outside the edges of the Cartesian plane comparable with Random Waypoint and Random Independent Movements has a behaviour comparable to the static model. All these tests were performed having $9$ edge nodes, placed in an optimized position (as shown in Figure\ref{pos}).

\subsection{Grid density}
\begin{figure}    
    \centering
    \includegraphics[width=.45\textwidth]{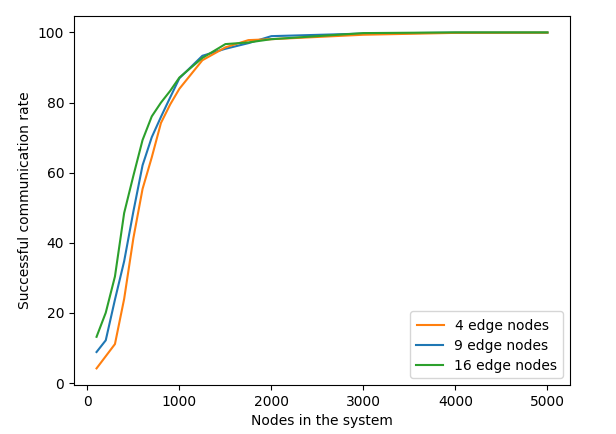}
    \caption{Successful communication rate achieved depending on the number of nodes on the grid.}
    \label{nodes}
\end{figure}

The approach that we propose to spread the information is assumed to work in a crowded environment, where there are several end-nodes that can contribute to the functioning of the system, by relaying the received messages. 
If the environment is scarcely populated, then there is the risk that either at some point the message gets lost or the number of hops in the routing process considerably grows. In Figure \ref{nodes}, we show how the successful communication range achieved changes by diminishing the population of the graph. In our default configuration, with {10\,000} nodes, $100\%$ of coverage is always achieved, and the same thing happens with more than $5000$ nodes. With a number of nodes ranging from $3000$ to $5000$, over 99.5\% of successful communication range is achieved. Then, when the amount of nodes is lower than $2000$, the coverage drops under $99\%$. In particular, the reliability of the communication starts to plummet with $1500$ nodes, where on average around $7.5$ nodes are expected to be found in the wireless radius.
Our experiments were carried out with a varying number of edge nodes, and pure broadcast was used in dissemination to guarantee that the maximum possible coverage is achieved.

\subsection{Edge nodes}

\begin{figure}    
    \centering
    \includegraphics[width=.45\textwidth]{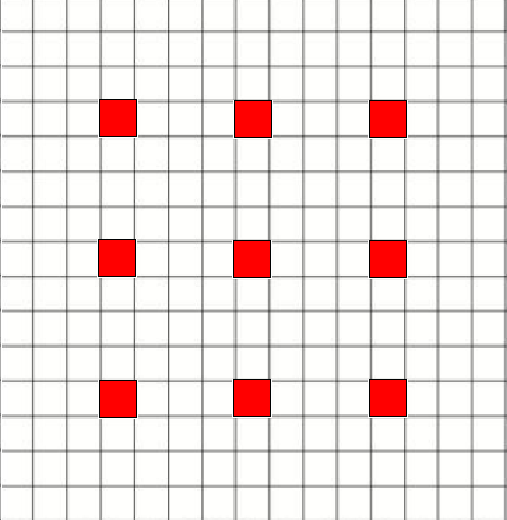}
    \caption{Optimal disposition of edge nodes (in red) in an empty grid.}
    \label{pos}
\end{figure}

\begin{figure}
    \centering
    \includegraphics[width=.45\textwidth]{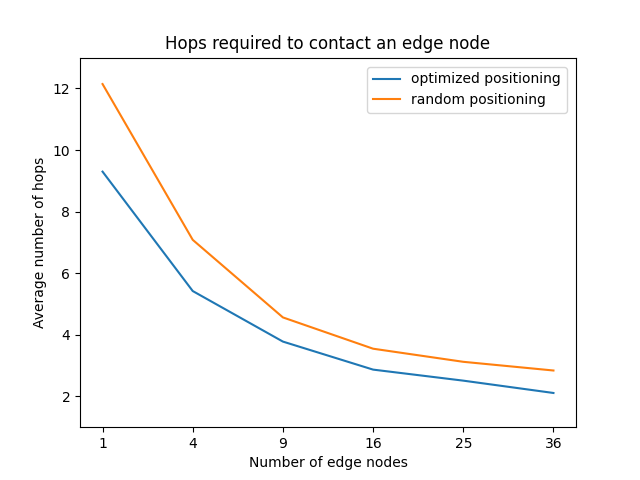}
    \caption{Average number of hops necessary to contact an edge node: comparison between an optimized positioning and a random positioning of the edge nodes.}
    \label{edgenodes}
\end{figure}

The amount of edge nodes and their arrangement around the Cartesian plane is of crucial importance for what concerns latency minimization. As easily predictable, the number of hops necessary to get in touch with an edge node is inversely proportional to the number of edge nodes, but also their geographical location has a significant impact on the metrics. 
Our testbed uses a simple environment, where there are neither obstacles nor areas characterized by a particular population density, and the agents are free to move along the grid. Therefore, the best way to optimize the position of the edge nodes is to follow the approach described by Figure~\ref{pos}, where the average distance between a random cell and a cell where there is an edge node is minimized.
Figure \ref{edgenodes} shows that with an optimized positioning, a lower number of hops is needed to contact an edge node. Despite the different results, the number of messages sent during the experiments was similar regardless of the configuration. This happens because the nodes are not informed about what happens during the dissemination, and therefore the messages propagation is not stopped when a destination is reached.

\subsection{Gossip protocols}
Pure broadcast is the most intuitive algorithm for message dissemination in P2P environments, and it ensures the best coverage and the minimum time delivery. However, a fine tuning of other protocols can lead to the same coverage with a significant number of messages saved, at the cost of a little growth on the average number of hops needed to reach an edge node.

\begin{figure}    
    \centering
    \includegraphics[width=.45\textwidth]{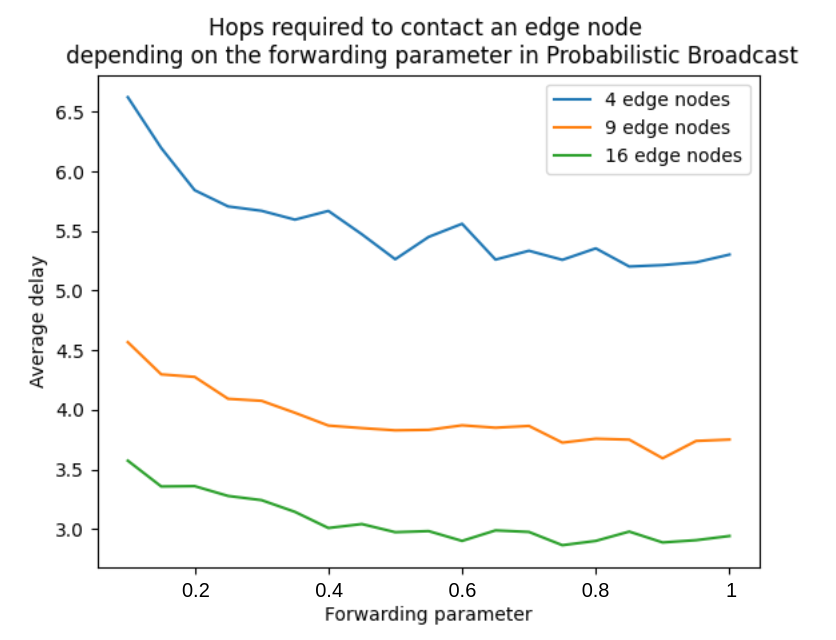}
    \caption{Average delay in Probabilistic Broadcast.}
    \label{PBdelay}
\end{figure}

\begin{figure}    
    \centering
    \includegraphics[width=.45\textwidth]{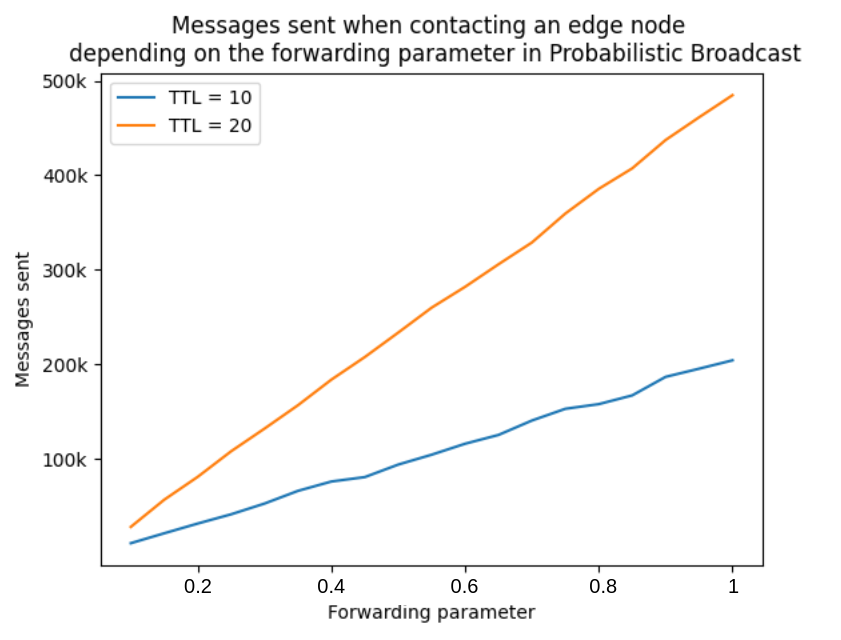}
    \caption{Average number of messages sent in Probabilistic Broadcast.}
    \label{PBmess}
\end{figure}

The following experiments aim at evaluating the performance of the algorithms and to understand which protocol ensures the best trade-off between time delivery and traffic minimization. We assume that for a correct functioning of the system, the successful communication rate achieved must tend towards 100\%, even though some applications could tolerate some level of messages loss\cite{yu2009loss}. 
The tests are performed with a varying number of edge nodes, placed in an optimized position. 
Figure~\ref{PBdelay} shows that, in Probabilistic Broadcast, the delay increases very slowly with the decrease of the forwarding parameter. On the other hand, the number of messages sent decreases quite linearly (Figure~\ref{PBmess}). If $0<p<1$ is the forwarding parameter, then the observed number of messages sent in our testbed is approximately $p * {500\,000}$. From what we observed, with $p > 0.4$, 100\% of successful communication range is always achieved, in the adopted simulation configuration.

Reducing the time-to-live from $20$ to $10$ hops allows to save a significant number of messages sent, at the cost of a slightly minor coverage achieved. In fact, almost always fewer than $10$ hops are needed to reach an edge node, but in some cases, above all when the forwarding parameter is low, a longer path is taken. The reduction in the number of delivered messages when time-to-live is reduced is due to the fact that nodes continue to propagate the information even if the destination is reached, because they cannot be informed if the communication has already been successful (or not).

\begin{figure}    
    \centering
    \includegraphics[width=.45\textwidth]{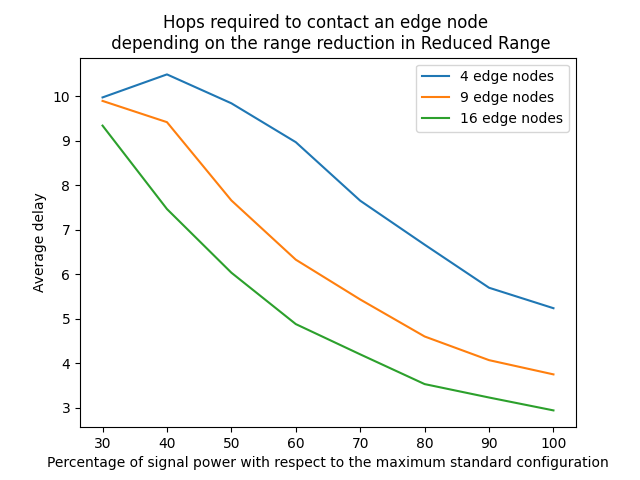}
    \caption{Average delay in Reduced Range.}
    \label{RRdelay}
\end{figure}

\begin{figure}    
    \centering
    \includegraphics[width=.45\textwidth]{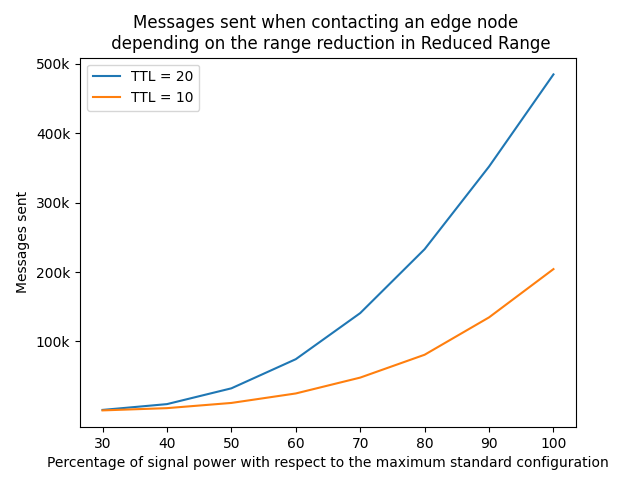}
    \caption{Average number of messages sent in Reduced Range.}
    \label{RRmess}
\end{figure}

Figure~\ref{RRdelay} shows that in the Reduced Range the delay sharply grows when the signal range is reduced. The saving of messages sent while ensuring 100\% of successful communication range (full coverage is observed with the signal reaching at least a distance of $25$ cell units) is high, but the cost in terms of time delivery due to the relay overhead may turn out to be significant. Figure~\ref{RRmess} shows that it is possible to save messages by reducing the time-to-live but, compared to Probabilistic Broadcast where the differential was minimal, in Reduced Range the decrease of the time-to-live leads to a significant reduction of the successful communication range.

\begin{figure}    
    \centering
    \includegraphics[width=.45\textwidth]{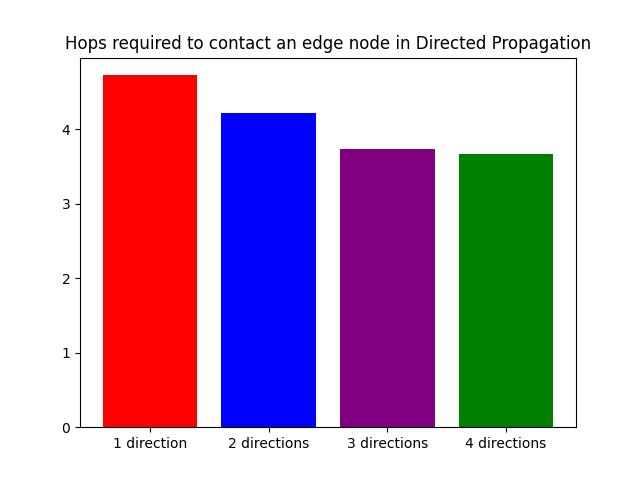}
    \caption{Average delay in Directed Propagation.}
    \label{DPdelay}
\end{figure}

\begin{figure}    
    \centering
    \includegraphics[width=.45\textwidth]{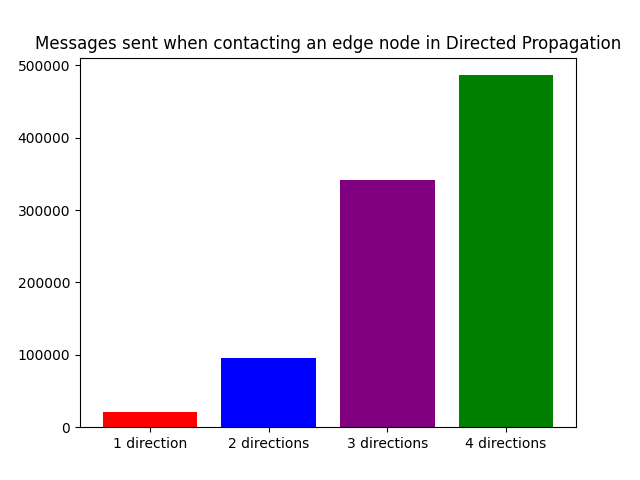}
    \caption{Average number of messages sent in Directed Propagation.}
    \label{DPmess}
\end{figure}

Finally, Figures~\ref{DPdelay} and \ref{DPmess} show that Directed Propagation is particularly efficient in terms of traffic minimization. In our experiments, we use geographical information to optimize the coverage (in all the four cases 100\% of successful communication range is achieved). In fact, nodes relay messages toward the center of the grid, where the edge nodes are located.



\section{Conclusions}
In this paper, we proposed a model for edge computing where communication is carried out according to the peer-to-peer principles, exploiting the presence of the mobile nodes for spreading the information though the network. This scheme offers an adaptive and decentralized solution for routing and traffic management and is thought for crowded and dynamic environments.

We made use of modelling and simulation to reproduce the communication mechanisms of the proposed system and to investigate how the design choices can influence the efficiency of the system, in terms of network traffic, time delivery and reliability of the communication. From a simulation point of view, it is interesting to observe that we used a multi-layer graph as a testbed for the experiments, where nodes are either edge nodes or end-users and connections are established by proximity, assuming that the emitted signals can propagate through a certain distance range.

Through simulation, we have demonstrated that the placement of the edge nodes and the employed dissemination strategy have a relevant impact on the metrics. More specifically, results show that it is helpful to place the edge nodes in strategic positions, in order to minimize their average distance with the users devices. Moreover, different dissemination protocols have been investigated and it turned out that certain strategies can lead to a consistent reduction of the messages being sent while not significantly worsening the reliability and the speed of the communication. In particular, exploiting geographical knowledge to direct the messages toward a certain location has turned out to be particularly efficient. The features of the various dissemination schemes might be combined, possibly taking into consideration environmental factors such as the device density.

\section*{Acknowledgments}
This work has received funding from the European Union’s Horizon 2020 research and innovation programme under the Marie Skłodowska-Curie International Training Network European Joint Doctorate grant agreement No 814177 Law, Science and Technology Joint Doctorate - Rights of Internet of Everything.

This research was also funded in part by the University of Urbino through the ``Bit4Food'' research project.

\bibliographystyle{IEEEtran}
\bibliography{biblio}

\begin{thebibliography}{10}
\providecommand{\url}[1]{#1}
\csname url@samestyle\endcsname
\providecommand{\newblock}{\relax}
\providecommand{\bibinfo}[2]{#2}
\providecommand{\BIBentrySTDinterwordspacing}{\spaceskip=0pt\relax}
\providecommand{\BIBentryALTinterwordstretchfactor}{4}
\providecommand{\BIBentryALTinterwordspacing}{\spaceskip=\fontdimen2\font plus
\BIBentryALTinterwordstretchfactor\fontdimen3\font minus
  \fontdimen4\font\relax}
\providecommand{\BIBforeignlanguage}[2]{{%
\expandafter\ifx\csname l@#1\endcsname\relax
\typeout{** WARNING: IEEEtran.bst: No hyphenation pattern has been}%
\typeout{** loaded for the language `#1'. Using the pattern for}%
\typeout{** the default language instead.}%
\else
\language=\csname l@#1\endcsname
\fi
#2}}
\providecommand{\BIBdecl}{\relax}
\BIBdecl

\bibitem{zichichi2020framework}
M.~Zichichi, S.~Ferretti, and G.~D’Angelo, ``A framework based on distributed
  ledger technologies for data management and services in intelligent
  transportation systems,'' \emph{IEEE Access}, pp. 100\,384--100\,402, 2020.

\bibitem{akpakwu2017survey}
G.~A. Akpakwu, B.~J. Silva, G.~P. Hancke, and A.~M. Abu-Mahfouz, ``A survey on
  5g networks for the internet of things: Communication technologies and
  challenges,'' \emph{IEEE access}, vol.~6, pp. 3619--3647, 2017.

\bibitem{ferretti2016smart}
S.~Ferretti and G.~D'Angelo, ``Smart shires: The revenge of countrysides,'' in
  \emph{2016 IEEE Symposium on Computers and Communication (ISCC)}.\hskip 1em
  plus 0.5em minus 0.4em\relax IEEE, 2016, pp. 756--759.

\bibitem{chourabi2012understanding}
H.~Chourabi, T.~Nam, S.~Walker, J.~R. Gil-Garcia, S.~Mellouli, K.~Nahon, T.~A.
  Pardo, and H.~J. Scholl, ``Understanding smart cities: An integrative
  framework,'' in \emph{2012 45th Hawaii international conference on system
  sciences}.\hskip 1em plus 0.5em minus 0.4em\relax IEEE, 2012, pp. 2289--2297.

\bibitem{kiss2018deployment}
P.~Kiss, A.~Reale, C.~J. Ferrari, and Z.~Istenes, ``Deployment of iot
  applications on 5g edge,'' in \emph{2018 IEEE International Conference on
  Future IoT Technologies (Future IoT)}.\hskip 1em plus 0.5em minus 0.4em\relax
  IEEE, 2018, pp. 1--9.

\bibitem{aslanpour2020performance}
M.~S. Aslanpour, S.~S. Gill, and A.~N. Toosi, ``Performance evaluation metrics
  for cloud, fog and edge computing: A review, taxonomy, benchmarks and
  standards for future research,'' \emph{Internet of Things}, p. 100273, 2020.

\bibitem{conti2018internet}
M.~Conti and A.~Passarella, ``The internet of people: A human and data-centric
  paradigm for the next generation internet,'' \emph{Computer Communications},
  vol. 131, pp. 51--65, 2018.

\bibitem{garcia2015edge}
P.~Garcia~Lopez, A.~Montresor, D.~Epema, A.~Datta, T.~Higashino, A.~Iamnitchi,
  M.~Barcellos, P.~Felber, and E.~Riviere, ``Edge-centric computing: Vision and
  challenges,'' 2015.

\bibitem{ccnc2020}
M.~{Zichichi}, S.~{Ferretti}, and G.~{D'Angelo}, ``A distributed ledger based
  infrastructure for smart transportation system and social good,'' in
  \emph{2020 IEEE 17th Annual Consumer Communications Networking Conference
  (CCNC)}, 2020, pp. 1--6.

\bibitem{gda-simpat-iot}
G.~D'Angelo, S.~Ferretti, and V.~Ghini, ``Multi-level simulation of internet of
  things on smart territories,'' \emph{Simulation Modelling Practice and Theory
  (SIMPAT)}, vol.~73, 2017.

\bibitem{Ferretti2013481}
S.~Ferretti, ``Shaping opportunistic networks,'' \emph{Computer
  Communications}, vol.~36, no.~5, pp. 481--503, 2013.

\bibitem{pads}
G.~D'Angelo, S.~Ferretti, and L.~Serena, ``{Parallel And Distributed Simulation
  (PADS) Research Group},'' \url{http://pads.cs.unibo.it}, 2021.

\bibitem{gda-ppna-2021}
L.~Serena, S.~Ferretti, and G.~D’Angelo, ``Cryptocurrencies activity as a
  complex network: Analysis of transactions graphs.'' \emph{Peer-to-Peer
  Networking and Applications}, 2021.

\bibitem{serena2020implications}
L.~Serena, G.~D'Angelo, and S.~Ferretti, ``Implications of dissemination
  strategies on the security of distributed ledgers,'' in \emph{Proceedings of
  the 3rd Workshop on Cryptocurrencies and Blockchains for Distributed
  Systems}, 2020, pp. 65--70.

\bibitem{backx2002comparison}
P.~Backx, T.~Wauters, B.~Dhoedt, and P.~Demeester, ``A comparison of
  peer-to-peer architectures,'' in \emph{Eurescom Summit}, vol.~2.\hskip 1em
  plus 0.5em minus 0.4em\relax Citeseer, 2002.

\bibitem{shi2016edge}
W.~Shi, J.~Cao, Q.~Zhang, Y.~Li, and L.~Xu, ``Edge computing: Vision and
  challenges,'' \emph{IEEE internet of things journal}, vol.~3, no.~5, pp.
  637--646, 2016.

\bibitem{varghese2016challenges}
B.~Varghese, N.~Wang, S.~Barbhuiya, P.~Kilpatrick, and D.~S. Nikolopoulos,
  ``Challenges and opportunities in edge computing,'' in \emph{2016 IEEE
  International Conference on Smart Cloud (SmartCloud)}.\hskip 1em plus 0.5em
  minus 0.4em\relax IEEE, 2016, pp. 20--26.

\bibitem{karagiannis2019edge}
V.~Karagiannis, A.~Venito, R.~Coelho, M.~Borkowski, and G.~Fohler, ``Edge
  computing with peer to peer interactions: Use cases and impact,'' in
  \emph{Proceedings of the Workshop on Fog Computing and the IoT}, 2019, pp.
  46--50.

\bibitem{yadgar2019modeling}
G.~Yadgar, O.~Kolosov, M.~F. Aktas, and E.~Soljanin, ``Modeling the edge:
  Peer-to-peer reincarnated,'' in \emph{2nd $\{$USENIX$\}$ Workshop on Hot
  Topics in Edge Computing (HotEdge 19)}, 2019.

\bibitem{ko2018wireless}
S.-W. Ko, K.~Han, and K.~Huang, ``Wireless networks for mobile edge computing:
  Spatial modeling and latency analysis,'' \emph{IEEE Transactions on Wireless
  Communications}, vol.~17, no.~8, pp. 5225--5240, 2018.

\bibitem{hyytia2007random}
E.~Hyyti{\"a} and J.~Virtamo, ``Random waypoint mobility model in cellular
  networks,'' \emph{Wireless Networks}, vol.~13, no.~2, pp. 177--188, 2007.

\bibitem{vastardis2014enhanced}
N.~Vastardis and K.~Yang, ``An enhanced community-based mobility model for
  distributed mobile social networks,'' \emph{Journal of Ambient Intelligence
  and Humanized Computing}, vol.~5, no.~1, pp. 65--75, 2014.

\bibitem{gda-jpdc-2017}
G.~D’Angelo and S.~Ferretti, ``Highly intensive data dissemination in complex
  networks,'' \emph{Journal of Parallel and Distributed Computing}, vol.~99,
  pp. 28 -- 50, 2017.

\bibitem{bononi2005scalable}
L.~Bononi, M.~Bracuto, G.~D'Angelo, and L.~Donatiello, ``Scalable and efficient
  parallel and distributed simulation of complex, dynamic and mobile systems,''
  in \emph{2005 Workshop on Techniques, Methodologies and Tools for Performance
  Evaluation of Complex Systems (FIRB-PERF'05)}.\hskip 1em plus 0.5em minus
  0.4em\relax IEEE, 2005, pp. 136--145.

\bibitem{gda-simpat-2017}
G.~D’Angelo, ``The simulation model partitioning problem: an adaptive
  solution based on self-clustering,'' \emph{Simulation Modelling Practice and
  Theory (SIMPAT)}, vol.~70, pp. 1 -- 20, 2017.

\bibitem{Ferretti2010}
S.~Ferretti and G.~D'Angelo, ``Multiplayer online games over scale-free
  networks: a viable solution?'' in \emph{Proc. of the 3rd International ICST
  Conference on Simulation Tools and Techniques}, ser. SIMUTools '10.\hskip 1em
  plus 0.5em minus 0.4em\relax Brussels, Belgium: ICST, 2010.

\bibitem{yu2009loss}
F.~Yu, V.~Gopalakrishnan, K.~Ramakrishnan, and D.~Lee, ``Loss-tolerant
  real-time content integrity validation for p2p video streaming,'' in
  \emph{2009 First International Communication Systems and Networks and
  Workshops}.\hskip 1em plus 0.5em minus 0.4em\relax IEEE, 2009, pp. 1--10.

\end{thebibliography}

\end{document}